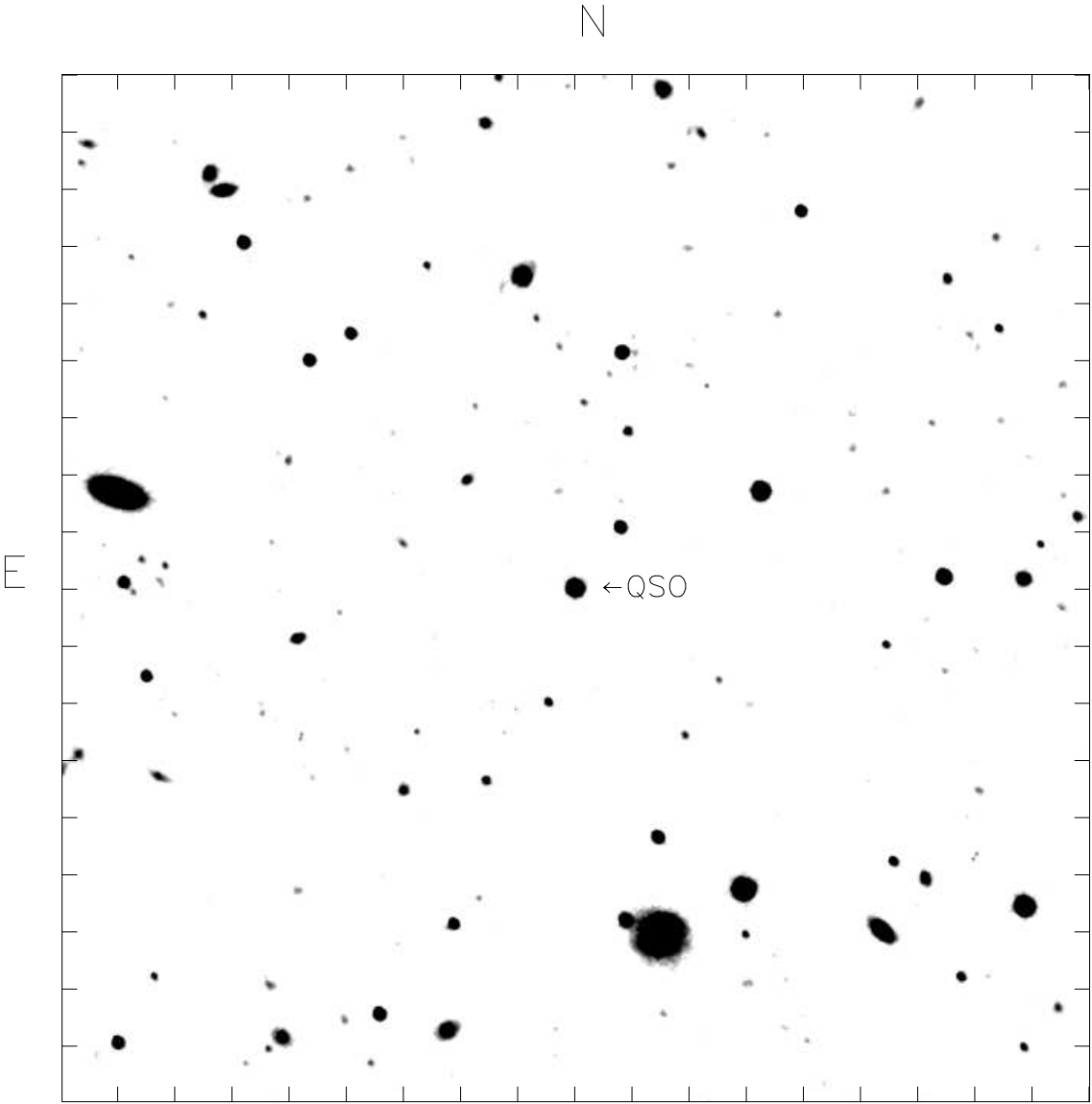

Figure 1

Figure 2

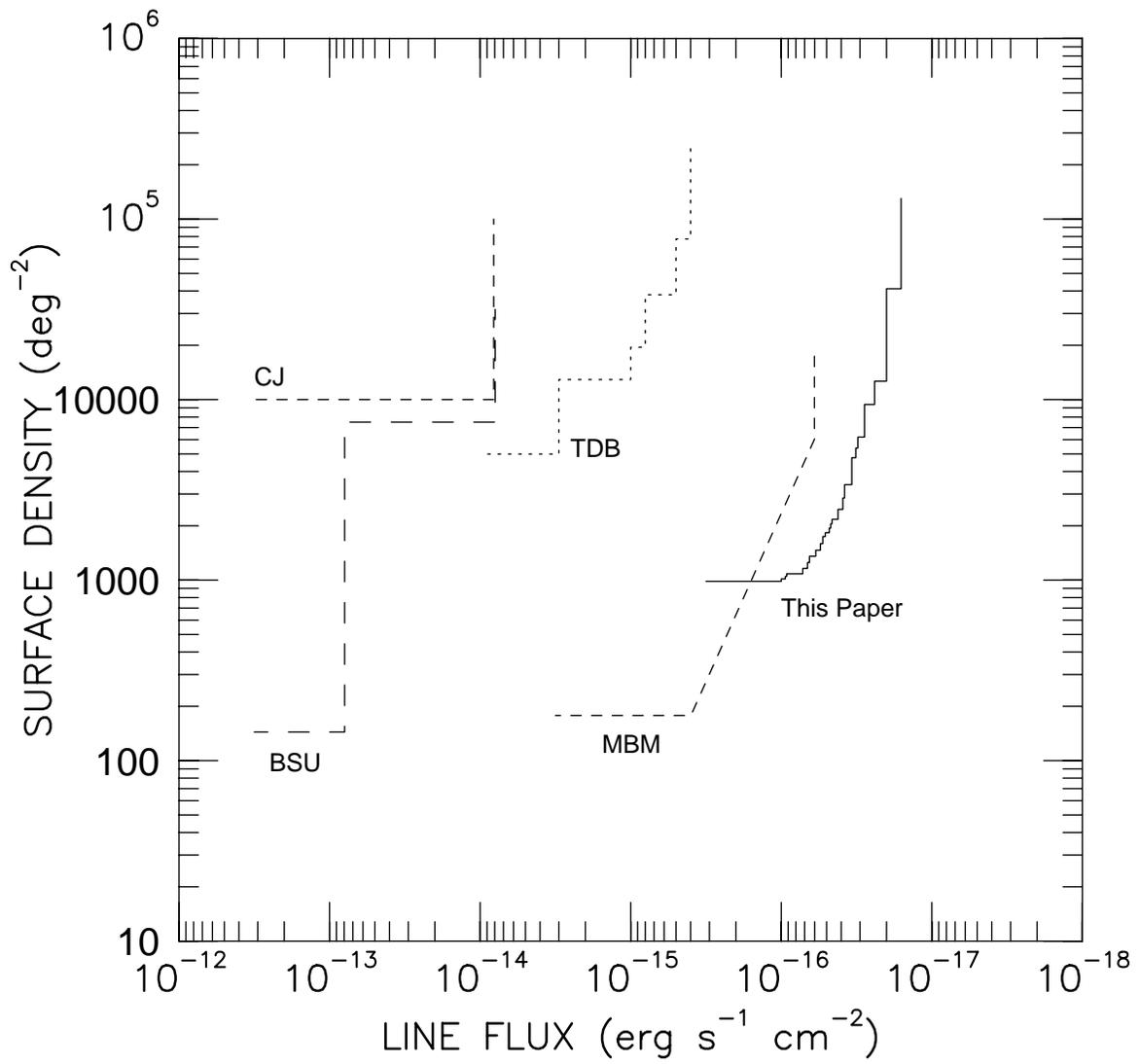

Figure 3

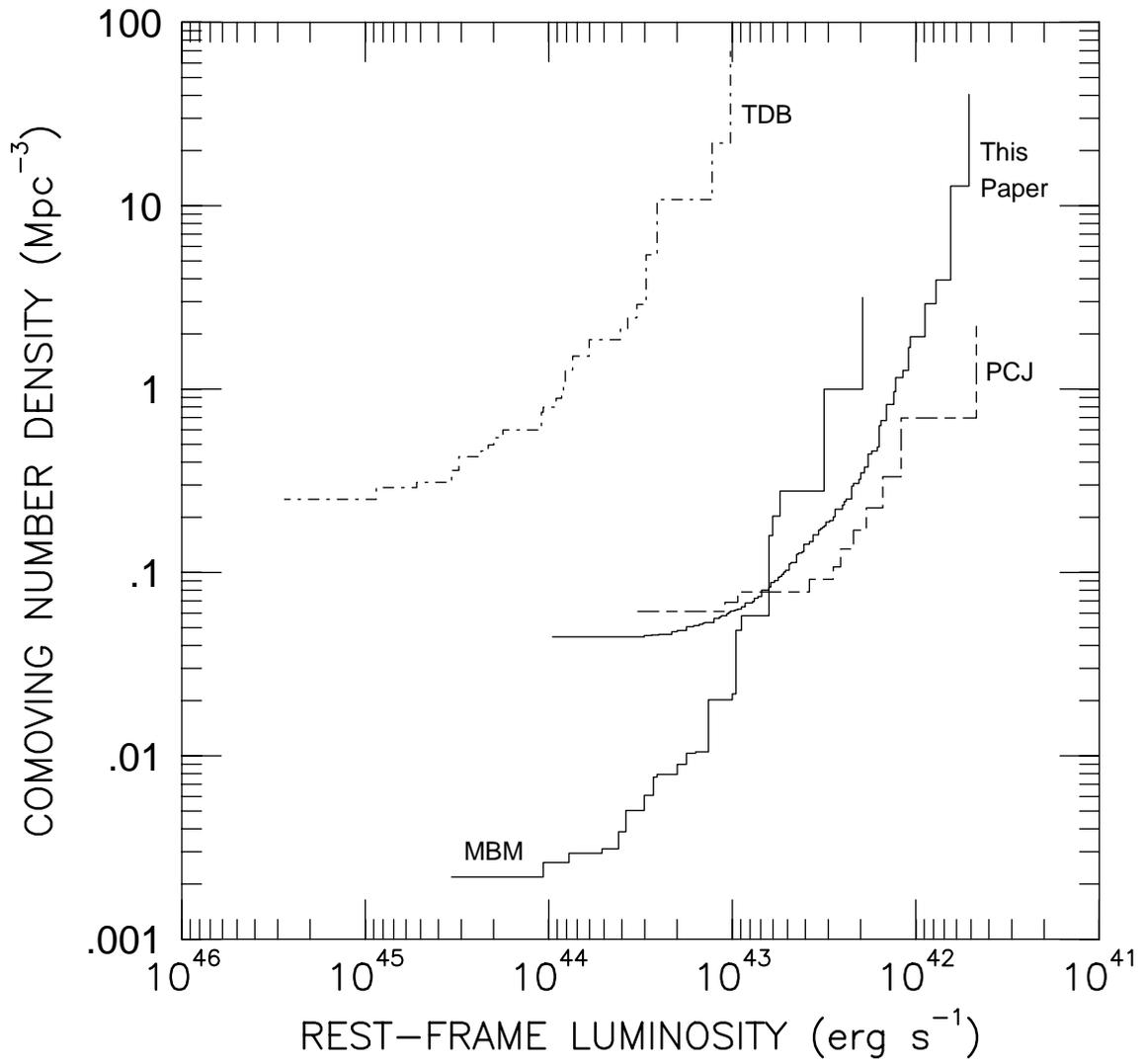

Figure 4

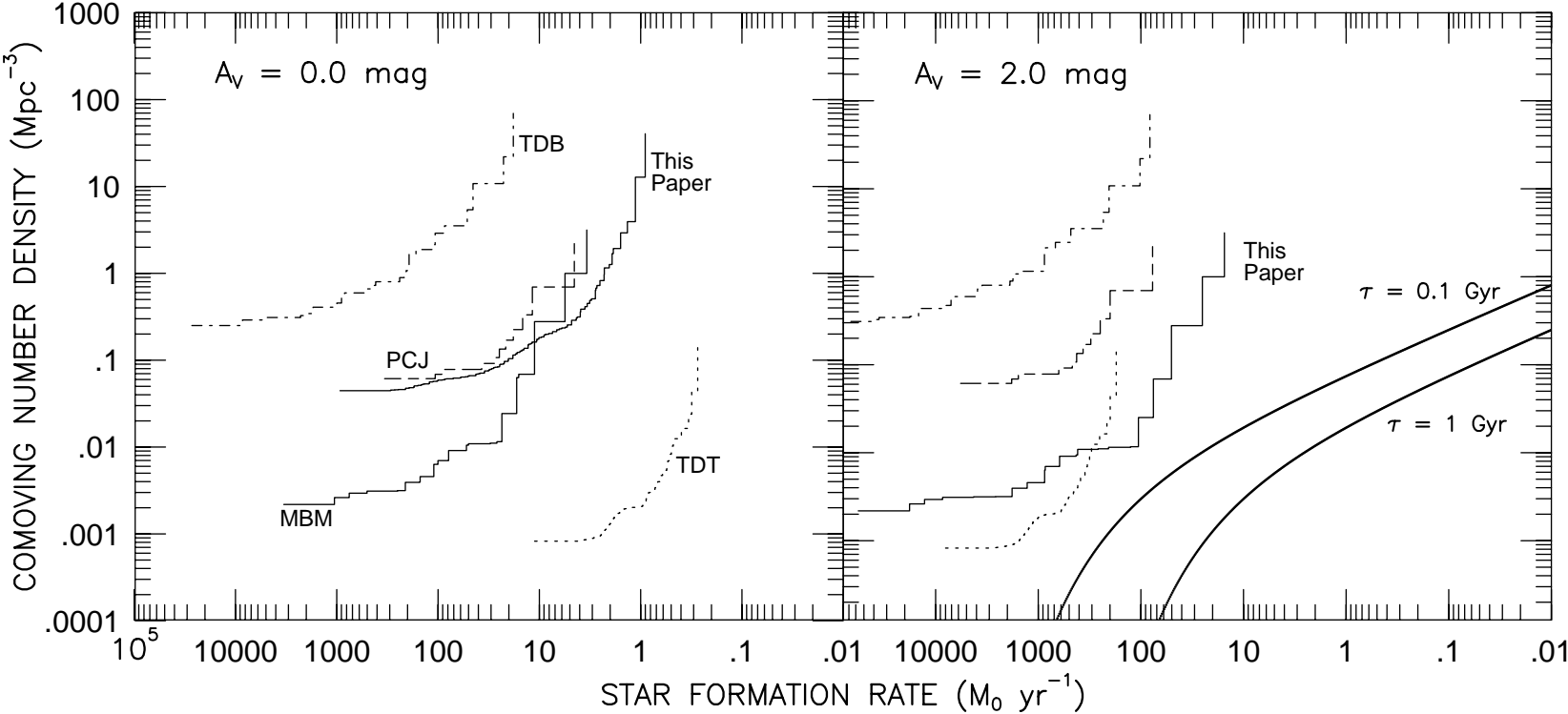

# A Near-Infrared Search for Line Emission from Protogalaxies Using the W.M. Keck Telescope[1]

Michael A. Pahre and S. G. Djorgovski[2]

Palomar Observatory, California Institute of Technology, Pasadena, CA 91125
*email*: map@astro.caltech.edu, george@oracle.caltech.edu

# ABSTRACT

We present the first results from a near-infrared narrow-band search for primeval galaxies (PGs) using the Keck 10 m telescope. We have targeted our fields for three QSOs and one radio galaxy at redshifts $z = 2.28$ to $4.70$. We selected narrow-band filters in the $K$-band centered on strong emission lines at the redshift of the targets. We reach limiting fluxes between $1.6 \times 10^{-17}$ erg s$^{-1}$ cm$^{-2}$ and $1.0 \times 10^{-16}$ erg s$^{-1}$ cm$^{-2}$, and cover a total area of $\sim 4$ arcmin$^2$. At the redshifts of interest, these flux limits correspond to typical restframe line luminosities of $\sim 10^{42} - 10^{43}$ erg s$^{-1}$, unobscured star formation rates of $\sim 1 - 100$ $M_\odot$ yr$^{-1}$, and a sampled comoving volume of several tens of Mpc$^3$. We have found no candidate PGs at a $2\sigma$ confidence level. We demonstrate that for moderate dust absorption ($A_V \gtrsim 2^m$) in a simple dust-screen model, our preliminary survey puts a strong constraint on intrinsic PG luminosities during an intial burst of star formation. In the case of the QSOs, we have used PSF-subtraction to search for the presence of faint, extended line emission surrounding these radio-quiet AGNs. We find no evidence for extended emission down to $3.7 \times 10^{-17}$ erg s$^{-1}$ cm$^{-2}$. This imposes limits on the reprocessed fraction of the QSO continuum emission ranging between 0.02% and 0.29%; if beaming effects are not important, then this implies similarly small covering factors for clouds optically thick to the QSO continuum emission.

*Subject headings:*

— Cosmology: Observations

— Galaxies: Formation

— Galaxies: Quasars: Emission lines

# 1. INTRODUCTION

Primeval galaxies (PGs), i.e., progenitors of present-day ellipticals and bulges undergoing their initial major bursts of star formation, have proved to be an elusive population despite thorough searches for their presence (for reviews see Djorgovski & Thompson 1992; Djorgovski 1992; or Pritchet 1994, and references therein). Most of the techniques search for the presence of narrow emission lines like Ly$\alpha$ as a signature (Partridge & Peebles 1967). Narrow-band searches for the Ly$\alpha$ emission line from PGs have sampled most of the relevant redshift space ($2 < z < 6$) down to very low limiting line fluxes over a large comoving volume but without finding viable candidate PGs; for a summary of limits, see, e.g., Djorgovski & Thompson 1992, Djorgovski, Thompson, & Smith (1993), or Thompson, Djorgovski, & Trauger (1995). The limits reached in those searches are probing at least an order of magnitude deeper than the predictions of the Baron & White (1987) CDM models, or of other simple models based on the local number density of galaxies.

A number of different explanations for this conspicuous absence of PGs have been proposed, including significant dust obscuration (with $A_V \sim 1 - 2^m$ required to bring the observations into agreement with the models; cf. Pritchet 1994, or Thompson *et al.* 1995), or an extended structure and a protracted star formation epoch for PGs in hierarchical merging scenarios. In the first explanation, PGs enshrouded in enough dust to be consistent with the observations may be in conflict with the observed absence of a sub-millimeter background measured with COBE (Mather *et al.* 1994; although see the discussion in Pritchet 1994). The latter explanation may be in conflict with the observed properties of elliptical galaxies, such as the mass-metallicity relation, high luminosity and phase space densities, and the existence of metallicity gradients.

If even a moderate amount of dust extinction is present, it may be worthwhile to expand the Ly$\alpha$ narrow-band searches to longer wavelength emission lines, such as [O II] (3727Å), H$\beta$ (4861Å), [O III] (5007Å), and H$\alpha$ (6563Å). In the galactic extinction models of Cardelli, Clayton, & Mathis (1989), $A_{Ly\alpha}/A_{H\alpha} = 4.28$. Within the redshift range of interest, $2 < z < 5$, all four of the above-mentioned lines are redshifted into the near-infrared $K$-band ($\lambda = 2.0$ to 2.4$\mu$m). Several groups have made preliminary searches for these emission lines in the $K$-band (Thompson, Djorgovski, & Beckwith 1994, hereafter TDB94; Mannucci, Beckwith, & McCaughrean 1994). While the initial results of the latter are approaching the surface density of PGs expected by simple models, neither has a very deep limiting flux.

Here we report on the results of a preliminary narrow-band imaging search for PGs in the near infrared, using the W.M. Keck telescope. A preliminary report of this investigation can be found in Pahre & Djorgovski (1994).

# 2. OBSERVATIONS AND DATA REDUCTIONS

We observed the fields around three high-redshift quasars, QSO 0856+406 (Djorgovski *et al.* 1995), QSO 1159+123 (Hazard *et al.* 1984), and BRI 1202−0725 (Isaak *et al.* 1994), and one high-redshift radio galaxy (RG), B3 0856+406 (Djorgovski *et al.* 1995), using the W.M. Keck 10-m telescope on Mauna Kea during the nights of UT 1994 April 5 and 6. The finding chart for BRI 1202−0725 is shown in Figure 1 (Plate ???). We have deliberately chosen to bias our sample towards the regions around known high-$z$ objects, which may mark the presence of clusters or groups at high redshifts. If so, our sample may be biased towards an overdensity of PGs. We note that two of our sources, QSO 0856+406 and B3 0856+406, are separated on the sky by only $\approx 37''$ (equivalent to 0.28 Mpc for $\Omega_0 = 0.2$ and $H_0 = 75$ km s$^{-1}$ Mpc$^{-1}$) and in redshift by $\approx 0.01$, and thus may in themselves signal the presence of a grouping at $z = 2.28$.

The observations were made using the near-infrared camera (NIRC; Matthews & Soifer 1994) operating within the $K$-band. The instrument is based on a $256 \times 256$ pixel In:Sb array, with $0.15''$ pixels, resulting in a full field-of-view of $38.4'' \times 38.4''$. We observed all four sources using the $K$ filter directly followed by observing the same source in a narrow-band filter. The narrow-band filters were selected from those normally available with the instrument so as to include one of the expected PG emission lines ([O II], H$\beta$, [O III], and H$\alpha$) redshifted into the filter. In all cases, we obtained grids of $3 \times 3$ pointings spaced by $10''$, with the total integration time of 1080 seconds per field per filter. We also observed standard stars from Casali & Hawarden (1992) in the $K$-band and all of the narrow-band filters except for the H$_2(1 \to 0)$ filter at $\lambda_0 = 2.1250 \mu$m, for which we assumed that the QSO has the same magnitude in both the broadband and narrow-band exposures.

## 3. A NARROW-BAND SEARCH FOR PG EMISSION LINES

Our final images have a full field-of-view of $57\rlap{.}''4 \times 57\rlap{.}''4$, and have sensitivities that vary across the field due to our $10''$ dithering procedure described above. The center of each field has a limiting flux listed in Table 1 over 315 arcsec$^2$, while the remaining portions of the field have relative limiting fluxes larger by a factor of 1.22 (710 arcsec$^2$), 1.5 (355 arcsec$^2$), 1.73 (710 arcsec$^2$), 2.12 (800 arcsec$^2$), and 3 (400 arcsec$^2$). For the typical seeing FWHM of $0\rlap{.}''75$, we calculate the $\sigma_{rms}$ for the noise in an optimized aperture (as measured in the simulations of Thompson et al. 1995) of diameter twice the seeing FWHM (i.e. $1\rlap{.}''5$). These limiting fluxes for each of the four fields are listed in the last column of Table 1. They are the deepest limiting fluxes for any $K$-band PG search yet attempted. In Figure 2 we compare our limiting fluxes and area-coverage to the $K$-band window narrow-band PG searches of TDB94 and Manucci et al. (1994), and to the $K$-band PG searches of Collins & Joseph (1988) and Boughn, Saulson, & Uson (1986). We have adopted these $1\sigma_{rms}$ limiting fluxes for the broadband surveys in the same way as by TDB94.

We have utilized two different methods to search our narrow-band images for PGs. We first identified all sources by eye and using FOCAS (Jarvis & Tyson 1981) using a $2\sigma_{rms}$ threshhold in both the narrow-band and broadband images, and compared lists to see if any sources appeared only in the narrow-band. The second method was to construct color-magnitude diagrams for the broadband and narrow-band images, in order to look for sources that brightened significantly ($\gtrsim 0.3^m$) in the narrow-band. For example, if we assume a constant $F_\nu$ from the Bruzual & Charlot (1993) models, then for the H$\alpha$ line we would expect a typical rest-frame equivalent width of order 100Å, which would result in a brightening of the source by $\sim 1^m$ in the narrow-band filter. To the limiting flux level of our four fields, we found no sources to have brightened in the narrow-band filter by $0.3^m$ at the $1\sigma$ confidence level, where $\sigma$ is taken to be the photometric errors.

Our survey can be considered to be a search for each of the four lines, H$\alpha$, [O III], H$\beta$, and [O II] at the redshifts appropriate for putting each of these four lines into each of the four narrow-band filters. We can transform our observed flux limits and redshift coverage for each line into a rest-frame luminosity and comoving volume sampled, using the approach of TDB94. For this purpose, we adopt a cosmology with $H_0 = 75$ km s$^{-1}$ Mpc$^{-1}$, $\Omega_0 = 0.2$, and $\Lambda_0 = 0$. The results are plotted in Figure 3. We have taken our best estimate for the flux limit and area coverage from Manucci et al. (1994), and have included the appropriate limits for [O II] at $z \approx 2$ for Parkes, Collins, & Joseph (1994).

These limiting rest-frame luminosities can be converted into limiting star formation rates (SFR), using the Kennicutt (1983) relation of $L(\mathrm{H}\alpha) = 5.0 \times 10^{40} h_{75}^{-2}$ erg s$^{-1}$ M$_\odot^{-1}$ yr. We scale the [O II], H$\beta$, and [O III] lines to H$\alpha$ using the line flux ratios of TDB94, and the Ly$\alpha$ to H$\alpha$ line ratio of 8.74 for the Case B recombination of Brockelhurst (1971). In the presence of dust, however, the H$\alpha$ line will undergo much less extinction than Ly$\alpha$, so we can calculate an equivalent limiting SFR by first scaling the rest-frame luminosity limits up by the extinction of the line sampled, according to the galactic extinction curves of Cardelli et al. (1989), and using a simple dust-screen model. The results are shown in Figure 4 for the unobscured case, and the dust screen models with $A_V = 0.0$ and $2.0^m$. We use for comparison a simple model in which the present mass of a galaxy formed in a time of 0.1 or 1 Gyr, and have utilized the local calibration of a Schechter luminosity function of $\phi* = 0.006 h_{75}^3$ Mpc$^{-3}$ and $M* = -19.2$ (Loveday et al. 1992), $\alpha = -1.5$, a mass-to-light ratio for spheroids of 6.6, and a fractional light contribution of spheroids of 1/3 (Schechter & Dressler 1987).

The optical searches for Ly$\alpha$ require extinction $A_V$ on the order of $2^m$ in order to keep the observational limits and the models of Baron & White (1987) in agreement (Pritchet 1994; Thompson et al. 1995). It is apparent from Fig. 4 that our $K$-band survey reaches the lowest SFR for $A_V \geq 2^m$, suggesting that we are beginning to sample the relevant regime of constraining the PG models in the presence of obscuration due to dust. These simple dust screen models underestimate the effects of extinction, relative to a more realistic situation where the dust, gas, and the photoionizing stars are spatially mixed. We note that the dust masses implied by the PG searches with this simple dust screen model may still be marginally allowed by the spectral distortions in the COBE fit to the CMB (Pritchet 1994).

## 4. CONSTRAINTS ON THE REPROCESSING OF QSO CONTINUUM RADIATION

High-redshift QSOs may be situated in young, gas-rich galaxies, and they might ionize the ISM of their host galaxies (Rees 1988). The search for low-surface brightness "fuzz" surrounding radio-loud AGN due to the reprocessing of continuum emission from the central QSO into line emission, has resulted in the discovery of a number of sources with either extended Ly$\alpha$ emission or Ly$\alpha$ emission companions (Djorgovski 1988; Djorgovski et al. 1987; Hu & Cowie 1987; Heckman et al. 1991ab; Hu et al. 1991). There do not seem to be, however, any detections of such line emission surrounding or in the vicinity of radio-quiet QSOs, with the possible exception of a companion source to Q1548+0917 (Steidel, Sargent, & Dickinson 1991), which may or may not be emitting reprocessed radiation. Our observations of three radio-quiet QSOs can help to further constrain the presence of such associated reprocessing clouds surrounding the central engines.

We have subtracted appropriately scaled broadband images of each QSO from the narrow-band images in order to look for extended line emission (the PSF widths are within 5%, so that subtraction artifacts should be minimal). We measured the residual flux in annuli surrounding the sources, and have concluded that in all three sources there is no detected extended emission. We can calculate the limit that this imposes on the percentage of reprocessed radiation from the QSO by integrating a power-law spectrum for the QSO with an assumed shape of $F_\nu \propto \nu^{-\alpha}$ from the Lyman continuum break to $\nu = \infty$, calibrate the spectrum from the $K$ broadband continuum flux, and measure the ratio of the limiting flux in the narrow-band filter to the total flux in the broadband filter. We assume $L_{Ly\alpha}/L_{H\alpha} = 10$, $L_{Ly\alpha}/L_{[O\ II]} = 33$, and $L_{Ly\alpha}/L_{[O\ III]} = 20$ from Osterbrock (1989) and Netzer (1990), and also that the reprocessed radiation is emitted isotropically, so that the lack of detected emission could also correspond to a limit on the covering factor of optically-thick clouds in this simple model. We assume $\alpha = 0.75$ as typical (Schneider, Schmidt, & Gunn 1991), and find that ranges of $0.5 < \alpha < 1.1$ produce less than a factor of two variation in the results in the highest-redshift case.

For radii between 1″ and 2″, we obtain the upper limits on the line fluxes of $8.0 \times 10^{-17}$ erg s$^{-1}$ cm$^{-2}$ for QSO 0856+406, $7.4 \times 10^{-17}$ erg s$^{-1}$ cm$^{-2}$ for QSO 1159+123, and $3.7 \times 10^{-17}$ erg s$^{-1}$ cm$^{-2}$ for BRI 1202–0725, from which we derive the upper limits on the fraction of the reprocessed QSO continuum of 0.18%, 0.024%, and 0.11%, respectively. For radii between 1″ and 3″, we obtain the flux limits of $1.3 \times 10^{-16}$ erg s$^{-1}$ cm$^{-2}$ for QSO 0856+406, $1.2 \times 10^{-16}$ erg s$^{-1}$ cm$^{-2}$ for QSO 1159+123, and $6.0 \times 10^{-17}$ erg s$^{-1}$ cm$^{-2}$ for BRI 1202–0725, and the corresponding limits on the fraction of the reprocessed continuum of 0.29%, 0.039%, and 0.18%, respectively. These limits are significantly stronger than previous estimates of Rees (1988) of $\lesssim 1\%$ reprocessed radiation, for the Ly$\alpha$ line. In our adopted cosmology of §3, a projected radius of 1″ corresponds to a scale of $\sim 7$ kpc for the redshifts of interest.

The associated equivalent widths implied by our limits are between 1 and 3.5Å in the observed frame. Such limits stand in direct contrast to the Ly$\alpha$ equivalent width of $\gtrsim 2000$Å for the Companion A to PKS 1614+051 (Hu & Cowie 1987; Djorgovski *et al.* 1987). It is important to note that our limits on extended emission are *not* for the Ly$\alpha$ line, but for the emission lines listed in Table 1 for each source which are then converted to Ly$\alpha$ limits using dereddened line intensity ratios. The non-detection of emission could be explained by a similarly small covering factor (as a fraction of $4\pi$ steradian) for optically-thick clouds, since we have assumed an isotropic radiation field for this simple model. If the QSO is enshrouded in a moderate quantity of dust, then we would expect limits on the longer-wavelength lines to be stronger constraints on the reprocessing of radiation, as has been argued above for the case of PGs. We note that if beaming effects are important in these QSOs, then it would be necessary to increase our limits on reprocessed radiation by the appropriate geometrical factor.

The W.M. Keck Observatory is a scientific partnership between the California Institute of Technology and the University of California. It was made possible by the generous and visionary gift of the W.M. Keck foundation, and the support of its president, Howard Keck. It is a pleasure to thank W. Harrison, T. Chleminiak, B. Schaefer, and W. Wack for the expert work during our observing run. We would also like to thank David Thompson for many helpful discussions. This work was supported in part by the NSF PYI award AST-9157412 to SGD, and the Greenstein and Kingsley Fellowships to MAP.

TABLE 1
SUMMARY OF OBSERVATIONS

| Source Name | $z$ | $\lambda_0$ ($\mu$m) | $\Delta\lambda$ ($\mu$m) | Targeted Line | Limiting Flux (erg s$^{-1}$ cm$^{-2}$) |
|---|---|---|---|---|---|
| QSO 0856+406  | 2.29  | 2.16   | 0.0224 | H$\alpha$  | $3.40 \times 10^{-17}$ |
| B3 0856+406   | 2.28  | 2.16   | 0.0224 | H$\alpha$  | $3.08 \times 10^{-17}$ |
| QSO 1159+123  | 3.502 | 2.2597 | 0.0531 | [O III]    | $3.15 \times 10^{-17}$ |
| BRI 1202-0725 | 4.695 | 2.1250 | 0.0235 | [O II]     | $1.60 \times 10^{-17}$ |

# FIGURE CAPTIONS

**Figure 1:** (Plate ???)   A $R_C$-band finding chart for the quasar BRI 1202–0725. The coordinates of the QSO, with an uncertainty of order $1''$, are: $\alpha(1950)=12^h02^m49\overset{s}{.}2$ and $\delta(1950)= -07°25'50''$. North is at the top and east to the left. The spacing between the tick marks on each axis is $10''$.

**Figure 2:**   Observed flux and surface density limits for $K$-band PG searches. The Collins & Joseph (1988; CJ) and Boughn, Saulson, & Uson (1986; BSU) surveys are broadband searches, while the Thompson, Djorgovski, & Beckwith (1994; TDB), Manucci, Beckwith, & McCaughrean (1994; MBM), and the present work are all narrow-band searches. The plotted limits for the flux are at the $1\sigma$ level in the same manner as in TDB94.

**Figure 3:**   Limiting comoving volume density and rest-frame luminosities for optical and near-infrared narrow-band PG searches for all possible emission lines for $1.5 < z < 9$. The labels are as in Fig. 2, with the addition of the Ly$\alpha$ surveys of Parkes, Collins, & Joseph (1994; PCJ), where we have also assumed that their $J$-band survey is sensitive to the [O II] line at $z \approx 2$, although not for the other emission lines (which lie at $z \approx 0.7$ to 1.3). For the transformation from observed to rest-frame parameters, we have assumed a cosmology of $H_0 = 75$ km s$^{-1}$ Mpc$^{-1}$, $\Omega_0 = 0.2$, and $\Lambda_0 = 0$.

**Figure 4:**   Limits on the star-formation rates of dusty PGs. We have plotted the limiting star-formation rate converted from the rest-frame luminosity as in Fig. 3. This has been done for two values of dust extinction ($A_V = 0.0$ and $2.0^m$) for a simple dust-screen model to demonstrate the effect of our limits on the presence of dust in PGs. Models using the local galaxy number density and a Schechter-type luminosity function are shown for formation times of 0.1 [upper] and 1 Gyr [bottom]. The cosmology is as assumed in Fig. 3. Dust absorption of $A_V \approx 2^m$ are required for the models and the observations to be in agreement. We have included the optical search for Ly$\alpha$ of Thompson *et al.* 1995 for comparison.